\documentstyle[aps,epsfig,multicol]{revtex} 
 
\begin{document} 
\title{Quantum phase transitions and magnetization of an integrable spin ladder 
with new parameters in bridging to real compounds } 
\author{Zu-Jian Ying$^{1,2,3}$, Itzhak Roditi$^1$, Huan-Qiang Zhou$^4$} 
\address{1. Centro Brasileiro de Pesquisas F\'\i sicas, Rua Dr. Xavier Sigaud 150, 
22290-180 Rio de Janeiro, RJ, Brasil\\ 
2. Department of Physics, Hangzhou Teachers College, Hangzhou 310012, China\\ 
3. Instituto de F\'\i sica da UFRGS, Av. Bento Gon\c calves, 9500, Porto 
Alegre, 91501-970, Brasil\\ 
4. Center for Mathematical Physics, School of Physical Sciences, The 
University of Queensland, 4072, Australia} 
\date{ } 
\maketitle 
 
\begin{abstract} 
We study the field-induced quantum phase transitions (QPT) and the relevant 
magnetic properties of a spin-1/2 two-leg integrable spin ladder (ISL), of 
which the system parameters in bridging to the real compounds are determined 
by setting the extra interactions in the Hamiltonian of the ISL relative to 
the Heisenberg spin ladder to vanish in the expectation in the ground state 
(GS). Such an ISL analytically has the correct leading terms of both the 
critical fields of the two QPT's as in the real strongly-coupled compounds:  
$g\mu _BH_{c1}=J_{\perp }-J_{\parallel }$ and $g\mu _BH_{c2}=J_{\perp 
}+2J_{\parallel }$ in terms of the experimental leg ($J_{\parallel }$) and 
rung ($J_{\perp }$) interactions. The symmetric magnetization inflection 
point is located at $g\mu _BH_{IP}=J_{\perp }+J_{\parallel }/2$. The 
magnetizations for the GS and at finite temperatures, as well as the 
susceptibility, show good agreements in various comparisons with the 
finite-site exact diagonalization, the transfer-matrix renormalization group 
numerical result, the perturbation theory, and the compounds  
(5IAP)$_2$CuBr$_4\cdot $2H$_2$O, Cu$_2$(C$_5$H$_{12}$N$_2$)$_2$Cl$_4$ and  
(C$_5$H$_{12}$N)$_2$CuBr$_4$. 
\end{abstract} 
 
%\pacs{75.10.Jm, 75.30.Kz, 75.40.Cx}  
 
\begin{multicols}{2} 
 
\section{Introduction} 
 
Recently spin ladders\cite{Dagotto} have attracted considerable interest due 
to increasing experimental realizations of ladder-structure compounds, such 
as SrCu$_2$O$_3$\cite{Azuma}, Cu$_2$(C$_5$H$_{12}$N$_2$)$_2$Cl$_4$\cite 
{Chaboussant}, (5IAP)$_2$CuBr$_4\cdot $2H$_2$O\cite{Landee},  
(C$_5$H$_{12}$N)$_2$CuBr$_4$\cite{Watson}, etc..  
Among the most interesting problems related 
to spin ladders are the existence of a gap, field-induced quantum phase 
transitions (QPT) and relevant magnetic properties. As it has been observed 
experimentally, strongly-coupled spin ladders are gapped and the magnetic 
field $H$ induces two QPT's with two critical fields. The first critical 
field $H_{c1}$ closes the gap and the second one $H_{c2}$ fully polarizes 
all the spins. In terms of the weak leg ($J_{\parallel }$) and strong rung  
($J_{\perp }$) interactions, the leading terms of the critical fields are  
$g\mu _BH_{c1}=J_{\perp }-J_{\parallel }$ and $g\mu _BH_{c2}=J_{\perp 
}+2J_{\parallel }$\cite{Barnes,Reigrotzki,Zheng,Chaboussant,Mila}, which 
coincide with the experiments. Many spin ladder compounds can be described 
by the standard Heisenberg spin ladder (HSL), though some other compounds 
possess spin frustration and contain diagonal leg interactions. 
Theoretically, properties in a magnetic field have been discussed by 
numerical methods such as exact diagonalization\cite{Hayward}, 
transfer-matrix renormalization group (TMRG) technique\cite{YuLu}, quantum 
Monte Carlo simulations\cite{Johnston}, as well as other methods for 
strong-coupling limit like perturbation expansions\cite{GuYu},  
bosonization\cite{Chitra} and mapping into XXZ Heisenberg  
chain\cite{Mila}. Recently 
some effort was exerted in bridging the integrable spin ladder (ISL)\cite 
{WangY} to real spin ladder compounds by introducing a rescaling parameter  
$\gamma $ in the leg interaction\cite{XiwenSU4,XiwenPRL}, the 
high-temperature expansion (HTE)\cite{Shiroishi,Tsuboi} was developed for 
the isotropic ISL\cite{XiwenPRL} and also extended to the anisotropic ISL 
with an XYZ rung interaction\cite{YingXYZ}. For a real spin ladder material, 
it is favorable to form rung states, 
since it is unfavorable for the weak leg 
interaction to take apart the strongly-coupled rung states. It happens that 
the ISL is based on the permutation of the formed rung states without 
breaking them. Therefore, if the system parameters are properly chosen, the 
ISL can provide some reasonable information, while from the ISL the 
analysis, for its characteristic features such as the ground state (GS), 
excitations and the gap, the QPT's, the magnetization inflection point (IP) 
and thermodynamical properties in finite temperatures, is quite convenient. 
The rescaling parameter $\gamma $ introduced in the ISL yields a gap $\Delta 
=J_{\perp }-4J_{\parallel }/\gamma $ and the two critical fields $g\mu 
_BH_{c1}=\Delta $, $g\mu _BH_{c2}=J_{\perp }+4J_{\parallel }/\gamma $\cite 
{XiwenSU4,XiwenPRL}. As far as the leading terms are concerned, the value  
$\gamma \approx 4$ for strong coupling fits the gap (and consequently 
$H_{c1})$ of the 
HSL\cite{XiwenSU4}, but the second critical field $g\mu _BH_{c2}=J_{\perp 
}+J_{\parallel }$ does not agree with the experiment-coinciding result $g\mu 
_BH_{c2}=J_{\perp }+2J_{\parallel }$\cite{Chaboussant,Mila}. 
Another point is that the magnetization symmetric
inflection point (IP) $H_{IP}=(H_{c1}+H_{c2})/2$ takes the value $J_{\perp
}/(g\mu _B)$ for all choices of $\gamma $, the deviation from the
experimental result $H_{IP}\approx (J_{\perp }+J_{\parallel }/2)/(g\mu _B)$
will become more considerable, when the leg interaction gets stronger.
A more appropriate way based on more physical origin for deciding the system
parameters is lacking and expected to lay a closer bridge of the ISL to the
real spin ladder compounds.
 
In the present paper, we shall consider an ISL where the system parameters 
are determined by the vanishing of the expectation value of the different 
interactions of the ISL relative to the HSL, for the GS in the gapped phase 
and the fully-polarized phase. We find that such an ISL analytically has the 
correct leading terms of the critical fields both at the first and the 
second QPT's: $g\mu _BH_{c1}=J_{\perp }-J_{\parallel }$ and $g\mu 
_BH_{c2}=J_{\perp }+2J_{\parallel }$. The magnetization IP is located at  
$g\mu _BH_{IP}=J_{\perp }+J_{\parallel }/2$.  
Our result of the magnetizations 
for the GS and at finite temperatures compare well both with numerical 
computation from the TMRG as well as the finite-spin exact diagonalization 
and with the experimental data for the compounds (5IAP)$_2$CuBr$_4\cdot  
$2H$_2$O, Cu$_2$(C$_5$H$_{12}$N$_2$)$_2$Cl$_4$ and  
(C$_5$H$_{12}$N)$_2$CuBr$_4$. 
The magnetic susceptibility coincides with the experimental data, and 
comparison with perturbation theory for strong coupling shows a good 
agreement. 
 
\section{Exact equivalence of HSL and the ISL for $H\geq H_{c2}$} 
 
Many spin ladder compounds are described by the HSL with the Hamiltonian  
\begin{eqnarray} 
&&{\cal H}_{HSL}={\cal H}_{\parallel }+{\cal H}_{\perp }+{\cal M},  \label{H} 
\\ 
{\cal H}_{\parallel } &=&J_{\parallel }\sum_{i=1}^L{\cal H}_{\parallel 
}^{i,i+1},\ {\cal H}_{\parallel }^{i,i+1}={\vec S}_i\cdot {\vec S}_{i+1}+ 
{\vec T}_i\cdot {\vec T}_{i+1} ,  \nonumber \\ 
{\cal H}_{\perp } &=&J_{\perp }\sum_i{\vec S}_i\cdot {\vec T}_i,\ {\cal M} 
=-g\mu _BH\sum_i\left( S_i^z+T_i^z\right) ,  \nonumber  
\end{eqnarray} 
where ${\vec S}$ and ${\vec T}$ are spin-1/2 operators for the two legs, the 
three parts ${\cal H}_{\parallel }$, ${\cal H}_{\perp }$ and ${\cal M}$ 
denote the leg interaction, rung interaction and the Zeeman energy, 
respectively. $J_{\parallel }$ and $J_{\perp }$ are the leg and rung 
coupling strengths. There are totally $L$ rungs and $H$ is the external 
magnetic field. 
 
Before considering for all values of the magnetic field, we shall first take 
a simple look at the fully-polarized phase and show that the HSL is 
equivalent to an ISL for $H\geq H_{c2}$. For convenience we label the rung 
singlet as $\left| 0_i\right\rangle =\left| \uparrow \downarrow -\downarrow 
\uparrow \right\rangle _i$ /$\sqrt{2}$ and the rung triplet as $\left| 
1_i\right\rangle =\left| \uparrow \uparrow \right\rangle _i$, $\left| 
2_i\right\rangle =\left| \uparrow \downarrow +\downarrow \uparrow 
\right\rangle _i$ /$\sqrt{2}$, $\left| 3_i\right\rangle =\left| \downarrow 
\downarrow \right\rangle _i$ where the subscript $i$ denotes the $i$'th 
rung. The rung interaction leads to an energy $-3J_{\perp }/4$ for the 
singlet and $J_{\perp }/4$ for the triplet. After $H_{c2}$ all spin are 
polarized so the ground state is $\Psi _{FP}=\left| 1_11_2\cdots 
1_L\right\rangle $ for a ladder with total $L$ rungs$,$ which is exactly the 
eigenstate of the HSL ${\cal H}_{HSL}$. 
 
When the field is lowered down to the critical point $H_{c2}$ from the 
fully-polarized phase, spins begin to flip down. More elementally the spin 
flips are linear combinations of the strongly-coupled rung states. Less 
excitations lead to lower energy, so one the-most-favorable state will enter 
the GS first, which indicates the occurrence of the QPT. Since the rung 
state $\left| 3\right\rangle $ has higher energy in the presence of the 
field, we only need to consider $\left| 0\right\rangle $ and $\left| 
2\right\rangle $ for the critical point. The lowest excitation can be 
trapped out from a subspace where there are at most one $\left| 
0\right\rangle $ and $\left| 2\right\rangle $, all possible neighbors 
collected are \{$\left| 1_i1_{i+1}\right\rangle $, $\left| 
1_i0_{i+1}\right\rangle $, $\left| 1_i2_{i+1}\right\rangle $, $\left| 
0_i2_{i+1}\right\rangle $ (and $\left| 0_i1_{i+1}\right\rangle $, $\left| 
2_i1_{i+1}\right\rangle $, $\left| 2_i0_{i+1}\right\rangle $)\}. For all 
these neighbors we exactly have an equivalent relation  
\begin{equation} 
{\cal H}_{\parallel }^{i,i+1}=\frac 12P_{i,i+1},  \label{HeqP} 
\end{equation} 
where $P_{i,i+1}\equiv (2{\vec S}_i\cdot {\vec S}_{i+1} 
+\frac 12)(2{\vec T}_i\cdot {\vec T}_{i+1}+\frac 12)$  
is the permutation operator. $P_{i,i+1}$ 
exchanges the rung states $\varphi _i$: $P_{i,i+1}\mid \varphi _i\varphi 
_{i+1}^{\prime }\rangle =\mid \varphi _i^{\prime }\varphi _{i+1}\rangle $. 
Consequently the Hamiltonian in this subspace can be rewritten as  
\[ 
{\cal H}_{HSL}=\sum_i\left( \frac{J_{\parallel }}2P_{i,i+1}+J_{\perp } 
{\vec S}_i\cdot {\vec T}_i-g\mu _BH(S_i^z+T_i^z)\right)  
\] 
which is actually an ISL\cite{WangY}. As there is at most one $\left| 
0\right\rangle $ and one $\left| 2\right\rangle $, the exact eigenenergy 
from the Bethe ansatz approach\cite{Sutherland} is  
\begin{eqnarray} 
E &=&E_{FP}+\left( g\mu _BH\right) \left( N_0+N_2\right)   \nonumber 
 \\ 
&&\ -\frac 12J_{\parallel }\sum_{j=1}^{N_0+N_2}\frac 1{\mu _j^2+\frac 14} 
-J_{\perp }N_0 ,  \label{E2} 
\end{eqnarray} 
where $E^{FP}=L\left( \frac 12J_{\parallel }+\frac 14J_{\perp }-g\mu 
_BH\right) $ is the fully-polarized eigenenergy, $N_0$ and $N_2$ (taking the 
value $0$ or $1$) are the numbers of states $\left| 0\right\rangle $ and  
$\left| 2\right\rangle $, respectively. 
 
Apparently both $\left| 0\right\rangle $ and $\left| 2\right\rangle $ are 
gapful if $g\mu _BH>\max \{J_{\perp }+2J_{\parallel },2J_{\parallel }\}$, 
and the GS is fully-polarized. The value $\mu =0$ corresponds to the 
critical field $g\mu _BH_{c2}=\max \{J_{\perp }+2J_{\parallel 
},2J_{\parallel }\}$, which reaches the same result deduced from the 
instability of spin wave spectrum in the ferromagnetic phase\cite 
{Chaboussant} and that obtained from mapping onto an XXZ Heisenberg chain 
\cite{Mila}. The expression of the saturation point is exact\cite 
{Chaboussant}, more exactly speaking, for an even number of rungs. The 
one-particle Bethe ansatz wavefunction corresponding to $\mu =0$ provides  
\begin{equation} 
\Psi _{c2}=\sum_{i=1}^L(-1)^i\left| 1_11_2\cdots b_i\cdots 1_L\right\rangle , 
\end{equation} 
where $b_i=0,2$ according to favorable energy. For antiferromagnetic spin 
ladder with positive $J_{\perp }$ and $J_{\parallel }$, $\left| 
2\right\rangle $ has higher excitation energy, then $b_i$ is the singlet  
$\left| 0\right\rangle $ and $g\mu _BH_{c2}=J_{\perp }+2J_{\parallel }$. The  
$\Psi _{c2}$ will be used in our further discussion. For an odd number of 
periodic rungs, the term $2J_{\parallel }$ in the critical expression is 
replaced by $\left( 1-\cos [(1-1/L)\pi ]\right) J_{\parallel }$, as $\mu =0$ 
is not the solution in periodic boundary conditions. But this size effect 
will completely disappear in thermodynamical limit. 
 
\section{ISL in bridging to the compounds in all field values} 
 
After the simple look at the fully-polarized phase in the previous section, 
we proceed to discuss all values of the field applied on an 
antiferromagnetic spin ladder. The relation ${\cal H}_{\parallel }^{i,i+1}= 
\frac 12P_{i,i+1}$ from (\ref{HeqP}) does not hold for neighbors $\left| 
0_i0_{i+1}\right\rangle $, $\left| 2_i2_{i+1}\right\rangle $ or $\left| 
1_i3_{i+1}\right\rangle $, though it is exact for all the other kinds of 
neighbor states. To find a model effective for all value of fields, we shall 
consider such an ISL with the difference in ${\cal H}_0$  
\begin{eqnarray} 
\ &&{\cal H}_{ISL}={\cal H}_0+{\cal H}_{\perp }+{\cal M},  \label{ISL} \\ 
{\cal H}_0 &=&J_{\parallel }\sum_i{\cal H}_0^{i,i+1},\ {\cal H}_0^{i,i+1}= 
{\cal H}_{\parallel }^{i,i+1}+{\cal H}_{extra}^{i,i+1} ,  \nonumber  
\end{eqnarray} 
where  
\begin{equation} 
{\cal H}_{extra}^{i,i+1}=\frac{P_{i,i+1}}\gamma -{\cal H}_{\parallel 
}^{i,i+1}+\frac \alpha 2 
( {\vec S}_i\cdot {\vec T}_i+{\vec S} 
  _{i+1}\cdot {\vec T}_{i+1}    )  
   +c.  \label{Hextra} 
\end{equation} 
The rung interaction part ${\cal H}_{\perp }$ and Zeeman term ${\cal M}$ are 
exactly the same as the HSL ${\cal H}_{HSL}$. If ${\cal H}_{extra}^{i,i+1}=0$ 
, the ${\cal H}_0^{i,i+1}$ will reduce to the leg interaction of the HSL. 
The extra part ${\cal H}_{extra}^{i,i+1}$ makes the model integrable, but 
the biquadratic exchange ${\vec S}_i\cdot {\vec S}_{i+1}{\vec T}_i\cdot  
{\vec T}_{i+1}$\cite{Nersesyan}  
in $P_{i,i+1}$ does not appear in the HSL, though it is as 
weak as the leg interaction. We set and balance the parameters $\gamma ,$  
$\alpha $ and $c$ to reduce its influence. To minimize its extra effect, we 
consider in such a way that it vanishes in expectation  
\begin{equation} 
\langle J_{\parallel }\sum_{i=1}^L{\cal H}_{extra}^{i,i+1}\rangle =0, 
\label{Heq0} 
\end{equation} 
as much as possible in the GS. Our consideration aims to make the model 
exactly soluble on one hand, while on the other hand, the extra effect is 
minimized as much as possible. 
 
More complicated balance terms such as ${S}_i^z{T}_i^z$, ${S}_i^z+{T}_i^z$ 
can be taken into account in ${\cal H}_{extra}^{i,i+1}$ but they break the 
isotropy. Here we consider the simplest case (\ref{Hextra}) which preserves 
the isotropy when it is giving quite good results. To see the integrability, 
the Hamiltonian can be rewritten as  
\begin{equation} 
{\cal H}_{ISL}=J_{\parallel }\sum_i(\frac{P_{i,i+1}}\gamma +c)+(J_{\perp 
}+\alpha J_{\parallel })\sum_i{\vec S}_i\cdot {\vec T}_i+{\cal M} 
\label{ISL2} 
\end{equation} 
where $P_{i,i+1}$ forms the permutation operator in the SU(4) integrable 
model\cite{LiSU4,LiSU4BA} and ${\cal M}$ is the Zeeman term as in the HSL. 
The model (\ref{ISL2}) essentially is an ISL\cite{WangY} which can be 
exactly solved by Bethe ansatz approach\cite{Sutherland}. The exact 
eigenenergy reads  
\begin{eqnarray} 
E = -\frac {J_{\parallel }}\gamma \sum_j^{M^{(1)}}\frac 1{\mu _j^{(1)2}+ 
\frac 14}+\sum_{k=0}^3E_kN_k+(\frac 12+c)LJ_{\parallel }, \label{E-BA} \\ 
E_0 =-\frac 34(J_{\perp }+\alpha J_{\parallel }),\ E_1=\frac 14(J_{\perp 
}+\alpha J_{\parallel })-g\mu _BH,  \nonumber \\ 
E_2 =\frac 14(J_{\perp }+\alpha J_{\parallel }),\ E_3=\frac 14(J_{\perp 
}+\alpha J_{\parallel })+g\mu _BH,  \nonumber 
\end{eqnarray} 
where $E_k$ is the single-rung energy of the singlet ($E_0$) and triplet ($ 
E_1$, $E_2$, $E_3$), $N_k$ is the total rung number in the corresponding 
rung states. The rung state with lowest $E_k$ is chosen as the reference 
state, $M^{(1)}$ is the total number of rungs occupied by the other rung 
states. The singlet is chosen as the reference state before the IP $g\mu 
_BH_{IP}=(J_{\perp }+\alpha J_{\parallel })$ (decided by $E_0=E_1$), while  
$\left| 1\right\rangle $ is the reference state thereafter. The gapped phase 
lies in the former case and the fully-polarized phase locates in the latter. 
 
The gapped phase of the ISL is composed of the singlet. The magnetic field 
lowers the energy of the triplet component $\left| 1\right\rangle $ which 
enters the GS when the gap is closed at the first QPT. Further increase of 
the field brings all the components of the singlet out of the GS and another 
gap opens at the second QPT (we refer the gapped phase to be at $H<H_{c1}$, 
though the fully-polarized phase is also gapped). Since both the gapped 
phase and the fully-polarized phase only consist of one rung state ($\left| 
0\right\rangle $ or $\left| 1\right\rangle $, respectively), which is chosen 
as the reference state, $M^{(1)}$ in these two phases corresponds to the 
number of the excitations to the other three components. Then from the 
eigenenergy (\ref{E-BA}), it is very easy to find the excitation gap and the 
two critical fields  
\begin{eqnarray} 
g\mu _BH_{c1} &=&(J_{\perp }+\alpha J_{\parallel })-\frac 4\gamma 
J_{\parallel },  \nonumber \\ 
g\mu _BH_{c2} &=&(J_{\perp }+\alpha J_{\parallel })+\frac 4\gamma 
J_{\parallel },  \label{Hc12} 
\end{eqnarray} 
which, respectively, closes the gap of the gapped phase and open the gap of 
the fully-polarized phase. The $4/\gamma $ terms come from the lowest energy 
in the excitation band, $\mu _j^{(1)}=0$ corresponds to the bottom of the 
energy band. The critical fields as well as the IP realize the relation  
$H_{IP}=\frac 12\left( H_{c1}+H_{c2}\right) $, which is the magnetization 
symmetric point and agrees with the experiments, as discussed in the 
following. 
 
Experimentally strongly-coupled spin ladder compounds have two similar 
QPT's, separating the application of the external magnetic field into three 
phases. As usually described by the HSL, the fully-polarized state $\Psi 
_{FP}$ exactly is the system GS after the second phase transition, while the 
singlet is the most favorable state and dominates overwhelmingly in the 
gapped phase before the occurrence of the first QPT for the 
antiferromagnetically strongly-coupled real spin ladder compounds. 
 
Among the total three phases, the extra-interaction effect of the ISL can be 
minimized explicitly in two phases, gapped phase and the fully-polarized 
phase. Fulfilling the relation (\ref{Heq0}) in these two phase gives rise to  
\begin{eqnarray} 
\frac 1\gamma -\frac 12+\frac 14\alpha +c &=&0,\ H>H_{c2};  \label{FPeqv} \\ 
\frac 1\gamma -\frac 34\alpha +c &=&0,\ H<H_{c1},  \label{singlet} 
\end{eqnarray} 
where the term $1/2$ in the former equation comes from the expectation of 
the HSL leg interaction, while in the latter equation this expectation 
vanishes in the singlet phase. The solution is direct to obtain and we yield 
to the limitations  
\begin{equation} 
\frac 1\gamma =\frac 38-c,\ \alpha =\frac 12.  \label{parameters} 
\end{equation} 
Under these parameters, the GS has the exact equivalence ${\cal H}_{HSL}= 
{\cal H}_{ISL}$ for the fully-polarized phase $H>H_{c2}$ and the expectation 
equivalence $\langle {\cal H}_{HSL}\rangle =\langle {\cal H}_{ISL}\rangle $ 
for the gapped singlet phase $H<H_{c1}$. Then the IP is determined from the 
definite $\alpha $ in (\ref{parameters})  
\begin{equation} 
g\mu _BH_{IP}=J_{\perp }+\frac 12J_{\parallel },  \label{IP} 
\end{equation} 
which, as it turns out in the following comparisons, produces well the 
experimental value and TMRG numerical result. Note here that this 
experiment-coinciding IP expression is an analytic result from our 
consideration, and instead of from empirical parameter choosing, the 
presence of the term $\frac 12J_{\parallel }$ comes from the only 
requirement on the extra-interaction-effect minimizing (\ref{Heq0}) of the 
above two phases. We shall decide more definitely the parameter $\gamma $ by 
considering the critical points of the QPT's. 
 
\subsection{Equivalence at the 2nd QPT} 
 
Since the exact analysis for the HSL is available at the second QPT as 
discussed in the previous section, we first consider such a case that the 
extra interactions of the ISL also vanish at the second critical point, in 
addition to the requirement on the gapped phase and fully-polarized phase. 
Although ${\cal H}_{extra}^{i,i+1}$ does not vanish for a single 
neighbor-site pairs $\left| 1_i0_{i+1}\right\rangle $ at $H_{c2}$, the 
action of these extra terms together can cancel each other. In fact, 
noticing the relation (\ref{HeqP}) for the rung-state neighbors $\left| 
1_i1_{i+1}\right\rangle $ and $\left| 1_i0_{i+1}\right\rangle $ involved at 
the second critical point, one can easily obtain  
\begin{equation} 
\sum_{i=1}^L{\cal H}_{extra}^{i,i+1}\Psi _{c2}=[(\frac 1\gamma -\frac 12+ 
\frac \alpha 4)(L-4)+cL]\Psi _{c2}. 
\end{equation} 
Setting the extra-interaction effect to vanish  
$\sum_{i=1}^L{\cal H}_{extra}^{i,i+1}\Psi _{c2}=0$,  
in addition to the limitations (\ref{parameters}) from  
the afore-discussed two phases, leads us to the definite 
parameters  
\begin{equation} 
\gamma =\frac 83,\ \alpha =\frac 12,\ c=0.  \label{c=0} 
\end{equation} 
Thus the relation (\ref{HeqP}) is replaced by ${\cal H}_{\parallel }^{i,i+1}= 
{\cal H}_0^{i,i+1}$for fully-polarized phase $H>H_{c2}$ and  
$\sum_i{\cal H}_{\parallel }^{i,i+1}=\sum_i{\cal H}_0^{i,i+1}$  
at the critical point $H_{c2} $. Such a simple consideration immediately  
gives the two critical fields  
\begin{eqnarray} 
g\mu _BH_{c1} &=&J_{\perp }-J_{\parallel },  \nonumber \\ 
g\mu _BH_{c2} &=&J_{\perp }+2J_{\parallel }. 
\end{eqnarray} 
As it is expected, $H_{c2}$ is the same exact result as we discuss in the 
previous section. Note that here $H_{c1}$ for the first QPT is actually 
composed of the leading terms from the perturbation theory and has been 
often used in experiments, while we have not imposed any limitation on the 
first QPT and it is a by-product of the consideration of the second QPT. 
 
\begin{figure}[p] 
\setlength\epsfxsize{70mm} 
\epsfbox{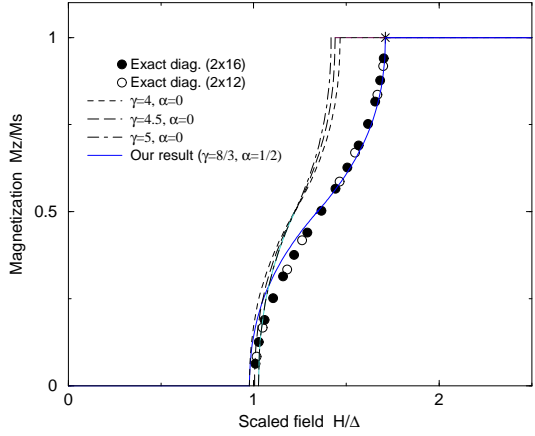} 
\caption{Magnetization versus the scaled magnetic field for the 
ground state (GS) (T=0.0K), $J_{\parallel }/J_{\perp }=0.2$. The symbols are 
result of Lanczos exact diagonalization on 2$\times $12 and 2$\times $16 
Heisenberg spin ladder (HSL) with periodic boundary conditions, the data are 
figured out from Ref.[10]. %%Ref.\cite{Hayward}.  
$\Delta $ is the gap of the HSL [28]. %%\cite{HaywardGap}.  
Our result (solid curve) from $\gamma =8/3$ and $\alpha =1/2$ 
coincides with that of the exact diagonalization, there is some small 
discrepancy at the first QPT due to existence of higher orders terms in  
$H_{c1}$. The dashed lines present result of some choices from an adjustable  
$\gamma $ and $\alpha =0$, they all go through a same half-saturation point 
away from the exact diagonalization data. 
} 
\label{compHayward} 
\end{figure}

\begin{figure}[p] 
\setlength\epsfxsize{70mm} 
\epsfbox{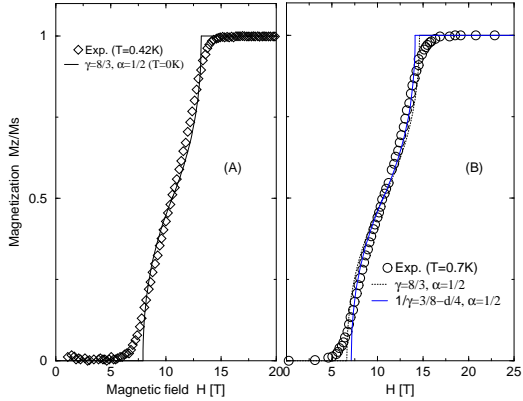} 
\caption{(A). Groundstate magnetization versus the magnetic field in 
comparison with the compound  
Cu$_2$(C$_5$H$_{12}$N$_2$)$_2$Cl$_4$ [3], %%\cite{Chaboussant},  
$g_z=2.03$, $J_{\perp }=13.2$K, $J_{\parallel }=2.4$K,  
$J_{\parallel }/J_{\perp }=0.18$. The symbols are experimental magnetization 
for the lowest available temperature (0.42K). The solid line is the 
groundstate (T=0.0K) magnetization calculated from our consideration with  
$\gamma =8/3$ and $\alpha =1/2$. 
(B). Groundstate magnetization compared with another compound  
(C$_5$H$_{12}$N)$_2$CuBr$_4$ [5], %%\cite{Watson},  
$g=2.13$, $J_{\perp }=13.3$K,  
$J_{\parallel}=3.8$K, $J_{\parallel }/J_{\perp }=0.29$. The temperature of the 
experimental data is 0.7K. The dotted line is the result of $\gamma =8/3$ 
and $\alpha =1/2$, while the solid line is the result including the high 
order terms in the gap, $1/\gamma =3/8-d/4$ and $\alpha =1/2$. The parameter  
$d$ is given by (\ref{JohnstonGap}) under certain values of $J_{\perp }$ and  
$J_{\parallel }$. 
} 
\label{MzGSexp} 
\end{figure}

\begin{figure}[p] 
\setlength\epsfxsize{70mm} 
\epsfbox{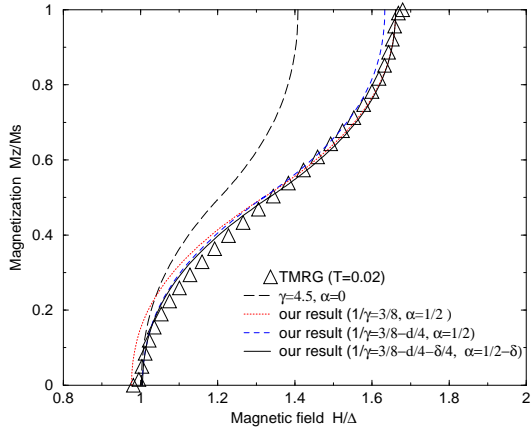} 
\caption{Groundstate magnetizations in different parameters in 
comparison with the T=0.02 TMRG result [11] %%\cite{YuLu},  
$J_{\perp }=$5.28,  
$J_{\parallel }=$1 and $\Delta =4.382$. Our reults are plotted according to  
(\ref{parameters}), (\ref{Hc1d}) and (\ref{paramHc12}), respectively, and  
$\delta =-d/2$ for the solid line. To show their small difference more 
clearly, we have focused the field on the region between 0.8 and 2.0.} 
\label{Hc12TMRG} 
\end{figure}

The magnetization can be obtained by thermodynamical Bethe ansatz (TBA)\cite 
{TBA,Frahm} which is also very convenient for analysis of the field-induced 
QPT's\cite{XiwenSU4,YingSzTz,YingXYZ}. For the ISL under consideration, the 
magnetization between the two critical fields in zero-temperature is decided 
by the following TBA equations of the dressed energy and density,  
\begin{eqnarray} 
\epsilon ^{(1)} &=&g^{(1)}-2\pi a_1-a_2*\epsilon ^{(1)-},  \nonumber \\ 
\rho ^{(1)} &=&a_1-a_2*\rho ^{(1)-},  \label{TBA} 
\end{eqnarray} 
where $g^{(1)}=\pm \gamma [(1+\alpha )J_{\perp }-g\mu _BH]/J_{\parallel }$ 
for $\pm (H_{IP}-H)>0$, $a_n(\mu )=[n/(\mu ^2+n^2/4)]/(2\pi )$, and the 
symbol $*$ denotes the convolution. The Fermi sea with negative dressed 
energy $\epsilon ^{(1)-}$ forms the system GS. Here the GS only involves one 
branch of dressed energy, since the GS of the ISL is composed of the singlet 
in the absence of the field, while only the component $\left| 1\right\rangle  
$ is lowered in energy level and brought down to the GS by the applied 
field. Correspondingly the gap of the excitation to the component $\left| 
2\right\rangle $ and $\left| 3\right\rangle $ never closes during the 
gapless competition between $\left| 0\right\rangle $ and $\left| 
1\right\rangle $ in the GS. We present an example of the GS magnetization in 
Fig.\ref{compHayward}. The symbols are data figured out from Ref.\cite 
{Hayward} in the study by Lanczos exact diagonalization on 2$\times $12 and  
2$\times $16 HSL with periodic boundary conditions, the ratio of the leg and 
rung interaction strength is $J_{\parallel }/J_{\perp }=0.2$. The solid 
curve is our result which coincides with the finite-site exact 
diagonalization. There is some small discrepancy at the first QPT due to 
existence of higher-order terms in $H_{c1}$ (the discrepancy will be reduced 
when the higher-order terms are picked up, as one will see from GS 
magnetization compared with T=0.02 TMRG result in Fig.\ref{YuLuMz} and  
Fig.\ref{Hc12TMRG}). As our consideration provides the same  
$H_{c1}$ and $H_{c2}$ 
as the HSL in leading terms, the coincidence of our result with the 
experiments is also expected. Two examples are given in Fig.\ref{MzGSexp}A 
and B, respectively for the compound  
Cu$_2$(C$_5$H$_{12}$N$_2$)$_2$Cl$_4$\cite{Chaboussant} with ratio  
$J_{\parallel }/J_{\perp }\approx 0.18$ and  
(C$_5$H$_{12}$N)$_2$CuBr$_4$\cite{Watson} with  
$J_{\parallel }/J_{\perp}\approx 0.29$. 
 
\begin{figure}[p] 
\setlength\epsfxsize{70mm} 
\epsfbox{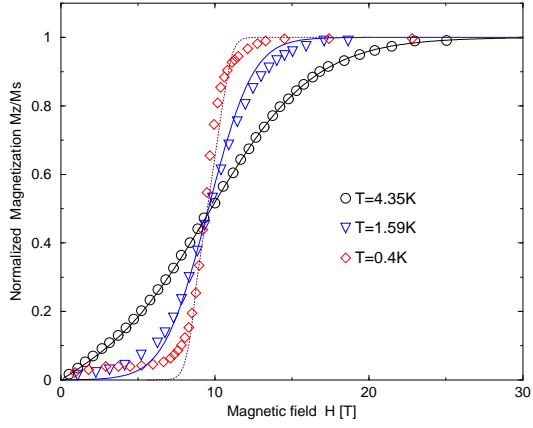} 
\caption{Magnetizations versus the magnetic field at different 
temperatures for the compounds (5IAP)$_2$CuBr$_4\cdot $2H$_2$O, $g=2.1$,  
$J_{\perp }=13.0$K, $J_{\parallel }=1.0$K. The data in symbols are the 
experimental result[4], %%\cite{Landee}, 
the lines are plotted from our theoretical 
calculation ($\gamma =8/3$ and $\alpha =1/2$). The compound contains a 
considerable amount of impurities, which can be seen obviously in the 
magnetization of T$=0.4$K case. The impurity effect is also reflected partly 
in the T=1.59K case but overwhelmed by higher temperature in T=4.35K. There 
is some uncertainty of the interactions ($J_{\parallel }/J_{\perp }=(7.7\pm 
1.5)\%$) due to the impurity. As the leg interaction $J_{\parallel }$ is 
rather small, the relative uncertainty is quite large. Here we use the 
values provided in the summary of Ref.[4]. %%\cite{Landee}. 
} 
\label{LandeeMz} 
\end{figure}

\begin{figure}[p] 
\setlength\epsfxsize{70mm} 
\epsfbox{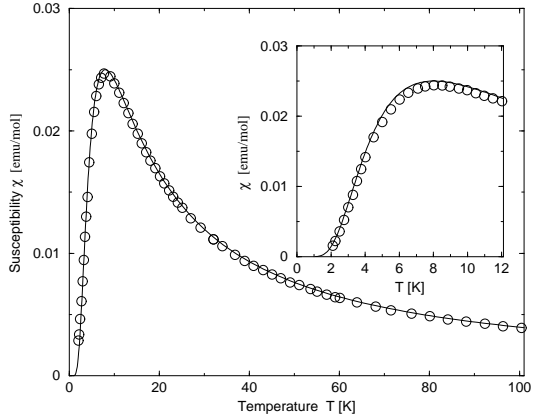} 
\caption{Magnetic susceptibility versus the temperature for the 
compounds (5IAP)$_2$CuBr$_4\cdot $2H$_2$O, $g=2.1$, the Curie constant  
$C=0.412$ emu K/mol, $J_{\perp }=13.0$K, $J_{\parallel }=1.0$K. The symbols 
are the experimental observation, the solid lines are from our calculation. 
The results for $\gamma =8/3$ and $1/\gamma =3/8-d/4$ do not differ, since 
the higher orders terms in the gap can be completely ignored due to the weak 
leg coupling ($J_{\parallel }/J_{\perp }\approx 0.077$). Impurity 
contribution was subtracted in the inset, so the experimental curve is 
lowered slightly. 
} 
\label{LandeeSus} 
\end{figure}

The high-temperature expansion (HTE)\cite{Shiroishi,Tsuboi} has been 
developed for the isotropic ISL\cite{XiwenPRL}. Then the 
temperature-dependent free energy per spin takes the form  
\begin{equation} 
f=-\frac 12T\sum_{n=0}^\infty C_n(\frac{J_{\parallel }}{\gamma k_BT})^n , 
\end{equation} 
where the expansion is carried out for per rung and the factor $1/2$ is 
added for per spin, $k_B$ is the Boltzmann constant. We present some orders 
of coefficients,  
\begin{eqnarray*} 
C_0 &=&\ln Q_{+},\ C_1=\frac{2Q}{Q_{+}^2},\ C_2=\frac{3Q}{Q_{+}^2}- 
\frac{6Q^2}{Q_{+}^4}+\frac{3Q_{-}}{Q_{+}^3}, \\ 
C_3 &=&\frac{10Q}{3Q_{+}^2}-\frac{18Q^2}{Q_{+}^4}+\frac{80Q^3}{3Q_{+}^6}+ 
\frac{8Q_{-}}{Q_{+}^3}-\frac{24QQ_{-}}{Q_{+}^5}+\frac 4{Q_{+}^4}, \\ 
C_4 &=&\frac{35Q}{12Q_{+}^2}-\frac{205Q^2}{6Q_{+}^4}+\frac{120Q^3}{Q_{+}^6}- 
\frac{140Q^4}{Q_{+}^8}+\frac{55Q_{-}}{4Q_{+}^3} \\ 
&&\ \  -\frac{100QQ_{-}}{Q_{+}^5}+\frac{180Q^2Q_{-}}{Q_{+}^7}- 
\frac{45Q_{-}^2}{2Q_{+}^6}+\frac{15}{Q_{+}^4}-\frac{40Q}{Q_{+}^6}, 
\end{eqnarray*} 
where  
\begin{eqnarray*} 
Q &=&2\cosh (\frac 12\beta \widetilde{J})+4\cosh (\frac 12\beta  
\widetilde{J} )\cosh (\beta h), \\ 
Q_{\pm } &=&2e^{(\pm \beta \widetilde{J}/4)}\cosh (\frac 12\beta  
\widetilde{J})+2e^{(\mp \beta \widetilde{J}/4)}\cosh (\beta h), 
\end{eqnarray*} 
$h=g\mu _BH$, $\widetilde{J}=(J_{\perp }+\alpha J_{\parallel })$ and $\beta 
=1/(k_BT$). One can get higher orders for lower temperatures. We compare our 
result of (\ref{c=0}) with the compound (5IAP)$_2$CuBr$_4\cdot $2H$_2$O\cite 
{Landee}, with ratio $J_{\parallel }/J_{\perp }\approx 0.077\pm 0.015$, the 
magnetizations $M=-\partial F/\partial H$ and the magnetic susceptibility  
$\chi =-\partial ^2F/\partial H^2$ are presented in Fig.\ref{LandeeMz} and 
Fig.\ref{LandeeSus}, respectively. The comparison shows a good agreement 
with the experiments. The unit of the susceptibility in comparing the 
experiment is determined by the Curie constant $C$, which can be clearly 
seen when we expand the susceptibility in (\ref{susN}). The Curie constant 
is decided, usually for molar susceptibility, when the $g$ factor is 
measured by the experiment. For the compound  
Cu$_2$(C$_5$H$_{12}$N$_2$)$_2$Cl$_4$\cite{Chaboussant},  
we have shown in Fig.\ref{MzGSexp}B the 
coincidence of our GS magnetization with the experimental magnetization of 
the lowest available temperature. This compound possesses nonlinear torque 
magnetization\cite{Crowell} and possible weak diagonal interaction\cite 
{Hayward}, which may be the possible reason that causes the observed 
asymmetric magnetizations at different higher temperatures. So here we shall 
not compare the magnetizations for this compound at higher temperatures. The 
compound (C$_5$H$_{12}$N)$_2$CuBr$_4$\cite{Watson} has stronger leg 
interaction ( $J_{\parallel }/J_{\perp }\sim 0.3$ ), the higher-order terms 
of the gap become more considerable. The small difference resulted from the 
high order terms can be seen from Fig.\ref{MzGSexp}B, the dotted line and 
the solid one are the magnetizations in the absence and presence of the 
high-order terms, respectively. In the following we shall pick up the 
high-order terms. 
 
\subsection{Fitting of the 1st QPT} 
 
\begin{figure}[p] 
\setlength\epsfxsize{70mm} 
\epsfbox{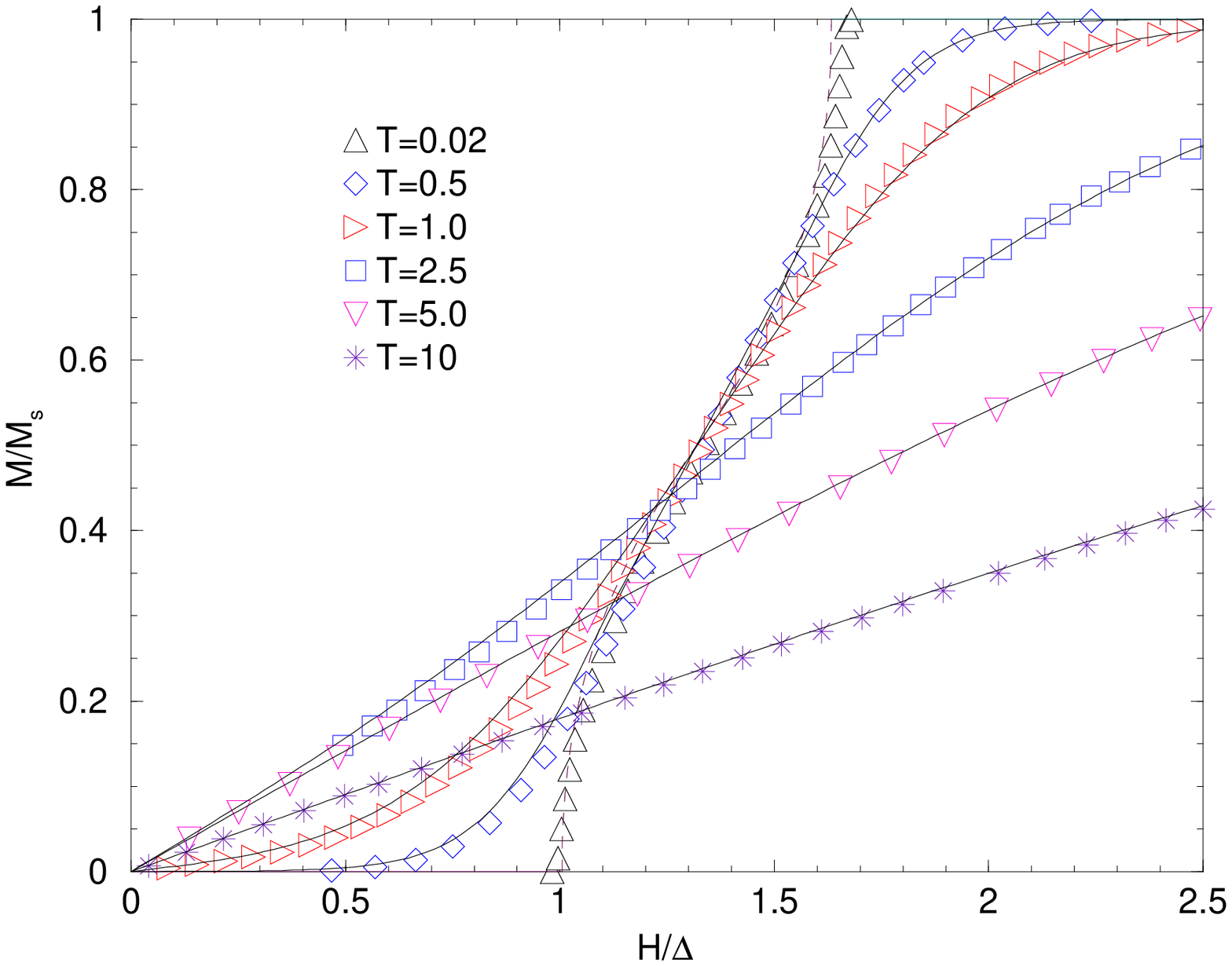} 
\caption{Magnetizations versus the magnetic field at different 
temperatures in comparison with the transfer-matrix renormalization group 
(TMRG) numerical result, $J_{\perp }=$5.28, $J_{\parallel }=$1 and $\Delta 
=4.382$. In this figure the $g$ factor and Bohr magneton $\mu _B$ are 
incorporated in the field $H$. The data in symbols are TMRG numerical  
result [11]. %%\cite{YuLu}.  
The solid lines and the broken line are plotted from  
our result for $1/\gamma =3/8-d/4$ and $\alpha =1/2$. The GS  
magnetization is plotted in the broken line in comparison  
with T=0.02 TMRG data.} 
\label{YuLuMz} 
\end{figure}

\begin{figure}[p] 
\setlength\epsfxsize{70mm} 
\epsfbox{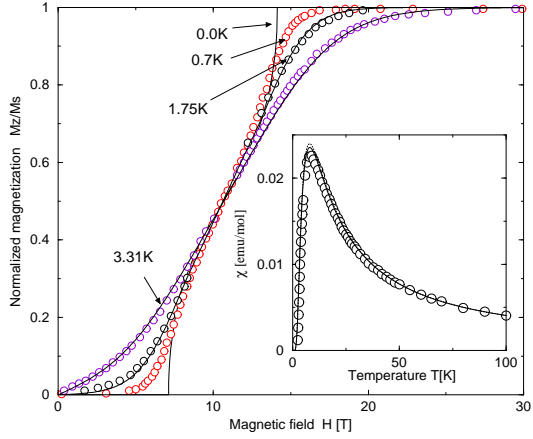} 
\caption{Magnetizations versus the magnetic field at different 
temperatures for the compound (C$_5$H$_{12}$N)$_2$CuBr$_4$, $g=2.13$,  
$J_{\perp }=13.3$K, $J_{\parallel }=3.8$K. The symbols denote the 
experimental data [5]. %%\cite{Watson}.  
The solid lines are plotted from our result 
for $1/\gamma =3/8-d/4$ and $\alpha =1/2$. The difference for the GS 
magnetization arising from the higher-order terms in the gap can be found in 
Fig.\ref{MzGSexp}B. The inset shows the susceptibility against the 
temperatures in a low field 0.1T, the Curie constant $C=0.425$ emu K/mol  
($g=2.13$) [5].%%\cite{Watson}.  
The solid line ($1/\gamma =3/8-(d+\delta )/4$,  
$\alpha =1/2-\delta $, $\delta =-d/2$) and dotted line ($1/\gamma =3/8-d/4$ 
and $\alpha =1/2$) have only small difference at the peak.} 
\label{WatsonMzSus} 
\end{figure}

Up to now our consideration has exhibited the correct leading terms of the 
critical fields. When the leg interaction is stronger, the higher-order 
terms in the gap (equal to $g\mu _BH_{c1}$) become more considerable. On the 
other hand, the fluctuation $\langle {\cal H}_{HSL}^2\rangle -\langle  
{\cal H}_{HSL}\rangle ^2$ in the singlet phase is very large, treatment of the 
first QPT may influence the system properties more sensitively than the 
second QPT. Unlike in the second QPT, the exact solution of the HSL 
is not available in the first QPT, requiring  
$\sum_i{\cal H}_{extra}^{i,i+1}\Psi _{c1}=0$  
is not explicitly executable. So we directly 
adjust the parameters to fit the gap, which is also a result of $\langle  
{\cal H}_{HSL}\rangle =\langle {\cal H}_{ISL}\rangle $. The higher orders 
terms in the gap of the HSL converge slowly in the 
perturbation theory, but it was found\cite{Johnston} that a two-parameter 
gap $\Delta _0$ fits the numerical data of Greven {\it et al}.\cite{Greven} 
very well  
\begin{equation} 
\Delta _0=J_{\perp }-J_{\parallel }+dJ_{\parallel },\ d=a(\frac{J_{\parallel 
}}{J_{\perp }})+b(\frac{J_{\parallel }}{J_{\perp }})^2,  \label{JohnstonGap} 
\end{equation} 
where $a=0.6878$, $b=-0.1861$. The $dJ_{\parallel }$ comes from 
the high-order terms in the gap and here we have used a label $d$ 
for our convenience. Setting  
\begin{equation} 
\frac 1\gamma =\frac 38-\frac d4,\ \alpha =\frac 12,\ c=\frac d4, 
\label{Hc1d} 
\end{equation} 
will give rise to this gap at the same time when the requirement in (\ref 
{parameters}) of the ISL is fulfilled. In Fig.\ref{YuLuMz} we compare the 
magnetizations with the TMRG numerical result\cite{YuLu} in various 
temperatures, for $J_{\parallel }=1$, $J_{\perp }=5.28$. The solid lines are 
plotted according to the corresponding temperatures of the TMRG data, the 
broken line is plotted at T$=0$ compared with the lowest temperature T=0.02 
TMRG result. As one can see, the agreement is quite good for all the 
temperatures. For the comparison with the experiment, Fig.\ref{WatsonMzSus} 
presents the magnetizations for the spin ladder compound  
(C$_5$H$_{12}$N)$_2$CuBr$_4$\cite{Watson}  
with a stronger leg interaction ($J_{\parallel}/J_{\perp }\approx 0.29$,  
$d\approx 0.18$). As the HTE is not valid for 
0.7K, we also calculate the magnetization of GS (0.0K) by TBA, a more clear 
comparison has been presented in \ref{MzGSexp}B. The figure demonstrates a 
good agreement of our result with the experiment. The susceptibility in a 
low field is given in the inset, our result in the solid line coincides with 
the experimental observation except for some small deviation at the hump. 
 
The interesting IP is an invariant point of the magnetizations at low 
temperatures. The compound (5IAP)$_2$CuBr$_4\cdot $2H$_2$O\cite{Landee} in 
Fig.\ref{LandeeMz} indicates the IP, though it suffers from the impurities. 
As the compound (C$_5$H$_{12}$N)$_2$CuBr$_4$ is free of the impurity or 
diagonal interaction, it exhibits an unambiguous IP at the half-saturation. 
The TMRG numerical calculation for the theoretically ideal HSL also shows 
the IP. From the ISL, it is direct to understand the IP\cite 
{YingXYZ,XiwenSU4,XiwenPRL}. Two reasons are both necessary for the IP: (i) 
equal energy of the only two components competing in the GS; (ii) large 
excitation gap relative to the temperatures. Actually only two components, 
the singlet $\left| 0\right\rangle $ and the triplet component $\left| 
1\right\rangle $, compete in the GS, as we mentioned below the dressed 
energy equation. The IP locates at the point where the reference state in an 
increasing field converts from $\left| 0\right\rangle $ to $\left| 
1\right\rangle $. At this point the two components have the same rung energy  
$E_0=E_1$ and same proportion $N_0=N_1$ in the GS. The excitations to the 
other components $\left| 2\right\rangle $ to $\left| 3\right\rangle $ are 
gapful, the gap are exactly available\cite{YingXYZ}  
\begin{equation} 
\Delta _{IP}=(J_{\perp }+\alpha J_{\parallel })-(2\ln 2)J_{\parallel 
}/\gamma ,  \label{gapIP} 
\end{equation} 
which is large due to the strong rung coupling $J_{\perp }$. Low 
temperatures can hardly stimulate excitations to $\left| 2\right\rangle $ or  
$\left| 3\right\rangle $, while each of $\left| 0\right\rangle $ and $\left| 
1\right\rangle $ with the same energy still occupies half the total rungs. 
As a result, the IP is invariant with a half-saturation magnetization under 
low temperatures. Higher temperatures will excite $\left| 2\right\rangle $ 
as well as $\left| 3\right\rangle $ and lower the magnetization, 
consequently the magnetization curves do not go through the IP, as one can 
see partly from the temperature T=2.5 and more clearly from T= 5.0, 10 in 
Fig.\ref{YuLuMz}. The location of the IP is well produced by our analytic 
expression (\ref{IP}) for the compound and the TMRG data. For the 
unambiguous IP of the compound (C$_5$H$_{12}$N)$_2$CuBr$_4$ ($g=2.13$,  
$J_{\perp }=13.3$K, $J_{\parallel }=3.8$K) our result $H_{IP}=10.6$T is 
identical with the experimental observation; the result for  
$J_{\perp }=$5.28, $J_{\parallel }=$1 from TMRG keeping 200  
optimal states provides an IP at $H_{IP}\approx 5.83$  
($H_{c1}\approx 4.38,H_{c2}\approx 7.28$ and $g\mu_B $ is incorporated  
in the field as in Ref.\cite{YuLu}) when our result is  
$H_{IP}=5.78$ with a very minor discrepancy about 0.9\%. 
 
\subsection{Fitting both the two critical fields} 
 
Under the parameters (\ref{Hc1d}) for fitting the gap, the extra 
interactions at the second critical field will be $J_{\parallel }\sum_{i=1}^L 
{\cal H}_{extra}^{i,i+1}\Psi _{c2}=-dJ_{\parallel }\Psi _{c2}$, which does 
not vanish though it is small. Nevertheless, this can be also resolved by 
more consideration. Previously we only take the dominant singlet component 
into account in the gapped phase of the HSL, as a simple physical picture as 
in the ISL. When the HSL-type leg interaction becomes less weak, small 
amount of other components will be mixed into the gapped singlet phase of 
the strong coupling limit. Therefore the average per-leg expectation value
of the HSL leg interaction does not vanish as in a pure singlet phase. This
contributes to another modification parameter $\delta $. As we 
are using the ISL to describe the compounds, the (\ref{singlet}) becomes  
\begin{equation} 
\frac 1\gamma -\frac 34\alpha +c=\delta ,\ H<H_{c1},  \label{mixed} 
\end{equation} 
instead. In limitation (\ref{singlet}), $\delta $ is null in our previous 
simpler consideration that singlet component dominates overwhelmingly in the 
strongly-coupled HSL. Combining (\ref{mixed}) in the gapped phase, the gap 
fitting $g\mu _BH_{c1}=\Delta _0$ and the extra-interaction minimizing (\ref 
{FPeqv}) in the fully-polarized phase, one gets  
\begin{equation} 
\frac 1\gamma =\frac 38-\frac{d+\delta }4,\ \alpha =\frac 12-\delta ,\ c= 
\frac \delta 2+\frac d4.  \label{paramHc12} 
\end{equation} 
If setting $\delta =-d/2$, one will regain the vanishing of the 
extra-interaction $\sum_{i=1}^L{\cal H}_{extra}^{i,i+1}\Psi _{c2}=0$ for the 
second critical field. In such case we will have the critical fields  
\begin{equation} 
g\mu _BH_{c1}=J_{\perp }-J_{\parallel }+dJ_{\parallel },\ g\mu 
_BH_{c2}=J_{\perp }+2J_{\parallel }, 
\end{equation} 
as well as a new IP  
\begin{equation} 
g\mu _BH_{IP}=J_{\perp }+\frac 12J_{\parallel }+\frac d2J_{\parallel }. 
\end{equation} 
For the compound (C$_5$H$_{12}$N)$_2$CuBr$_4$\cite{Watson}, these results 
does not give improvement to the magnetizations except some small 
improvement at the peak of the low-field susceptibility (given in the inset 
of the Fig.\ref{WatsonMzSus}). But for the theoretically ideal HSL under the 
couplings $J_{\perp }=5.28$, $J_{\parallel }=1$, the value $d\approx 0.124$ 
from (\ref{JohnstonGap}) gives the IP $g\mu _BH_{IP}=5.84$, which minimizes 
the discrepancy from the TMRG result $g\mu _BH_{IP}=5.83$\cite{YuLu}. As our 
previous considerations with simple parameters have given quite good 
results, here we just give a comparison for the GS magnetization with the 
TMRG data in Fig.\ref{Hc12TMRG}. 
 
\subsection{Susceptibility compared with perturbation theory of HSL} 
 
In bridging to the experimental data, the unit is a problem needing some 
special attention. The unit of the susceptibility can be clarified if the 
expansion is carried out. One can obtain the expansions of the 
susceptibility for $N$ spins easily from the free energy expression, the 
lowest two orders take the form  
\begin{equation} 
\chi =\frac CT\left[ 1-\frac 1{4k_BT}\left( J_{\perp }+(\frac 2\gamma 
+\alpha )J_{\parallel }\right) +{\cal O}(\frac 1{T^2})\right]  \label{susN} 
\end{equation} 
where $C=Ng^2\mu _B^2/(4k_B)$ is the Curie constant. In the expression  
(\ref{susN}) the part in the square brackets is dimensionless. The Curie  
constant is provided by the experiments corresponding 
to a measured value of $g$, e.g., $C=0.412$ emu K/mol for  
the compound (5IAP)$_2$CuBr$_4\cdot $2H$_2$O ($g=2.1$)\cite{Landee}.  
As one knows, the lowest 
order $\chi =C/T$ is the Curie law, which guarantees the coincidence with 
the experimental data in high temperatures ($T\rightarrow 100$K for the 
compounds discussed in the present paper). Comparing (\ref{susN}) with 
the Curie-Weiss Law $\chi =C/(T-\Theta )$ 
implies a Curie-Weiss temperature $\Theta _{ISL}=-[J_{\perp }+(2/\gamma 
+\alpha )J_{\parallel }]/4$ for the ISL (\ref{ISL2}). The HSL has a 
Curie-Weiss temperature $\Theta _{HSL}=-(J_{\perp }+2J_{\parallel })/4$\cite 
{Johnston}. Our parameters $\gamma =8/3,\ \alpha =1/2$ provide $\Theta 
_{ISL}=-(J_{\perp }+5J_{\parallel }/4)/4$, which is close to that of the 
HSL. The difference of $\Theta _{ISL}$ from $\Theta _{HSL}$ will be more 
ignorable when the leg interaction $J_{\parallel }$ is relatively small in a 
strongly-coupled spin ladder, which coincides with the good agreement of 
susceptibility for the compound (5IAP)$_2$CuBr$_4\cdot $2H$_2$O. 
 
\begin{figure}[p] 
\setlength\epsfxsize{70mm} 
\epsfbox{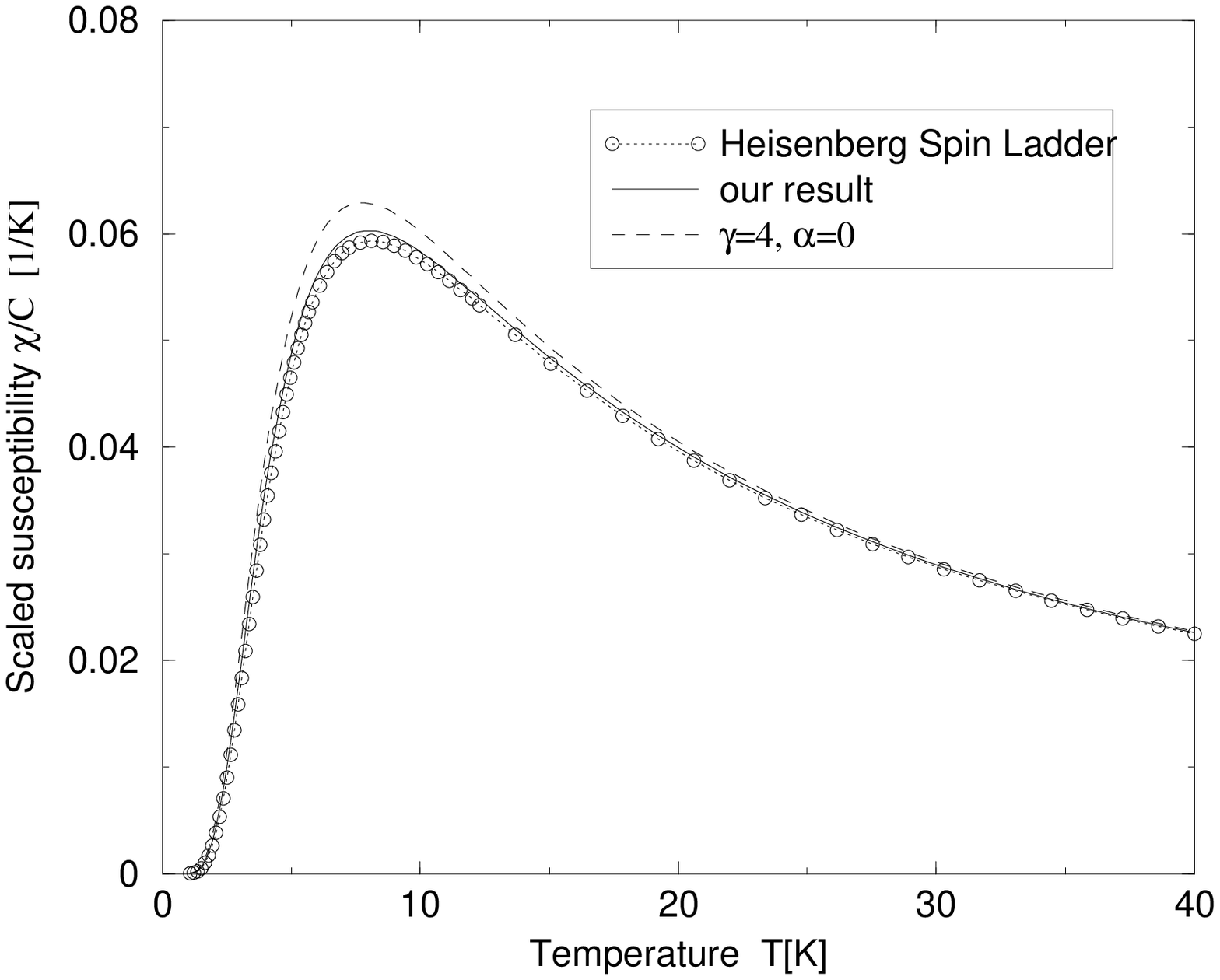} 
\caption{Scaled magnetic susceptibility against the temperature in 
comparison with the Heisenberg spin ladder (HSL), $J_{\parallel }/J_{\perp 
}=0.1$, $J_{\perp }=13$K. Since all compounds used in the figures  
have rung interaction around 
13K, we use this value here. For more clarity of the unit, we scale the 
susceptibility by the Curie constant $C$. The HSL curve is plotted from the 
expression (\ref{GuYuHSL}), which reproduces accurately the quantum Monte 
Carlo simulations [12] %%\cite{Johnston}  
 for $J_{\parallel }/J_{\perp }\leq 0.1$. 
Here the high-order terms in the gap make little difference in our result 
due to the small ratio $J_{\parallel }/J_{\perp }$. The agreement for our 
result with the HSL for stronger coupling $J_{\parallel }/J_{\perp }<0.1$ is 
even better. The broken line is the result from the choice $\gamma =4,\ 
\alpha =0$, for comparison. 
} 
\label{comp0.1} 
\end{figure}

For the strongly-coupled HSL, it was found that the expression of zero-field 
susceptibility derived from perturbation theory\cite{Johnston,GuYu},  
\begin{eqnarray} 
\chi &=&\frac{4C}T\left\{ \frac 1{3+e^{{\cal B}}}-\frac{J_{\parallel }} 
{J_{\perp }}\left[ \frac{2{\cal B}}{(3+e^{{\cal B}})^2}\right] \right.  
\nonumber \\ 
&&\ \ \ \ \ -\left( \frac{J_{\parallel }}{J_{\perp }}\right) ^2\left[ \frac{3 
{\cal B}(e^{2{\cal B}}-1)-{\cal B}^2(5+e^{2{\cal B}})}{4(3+e^{{\cal B}})^3} 
\right]  \nonumber \\ 
&&\ \ \ \ \ -\left( \frac{J_{\parallel }}{J_{\perp }}\right) ^3\left[ \frac{3 
{\cal B}(e^{2{\cal B}}-1)}{8(3+e^{{\cal B}})^3}\right.  \nonumber \\ 
&& \ \left. \left. -\frac{9{\cal B}^2e^{{\cal B}}(1+3e^{{\cal B}})-{\cal B 
}^3(7e^{2{\cal B}}-9e^{{\cal B}}-12)}{12(3+e^{{\cal B}})^4}\right] \right\}, 
\label{GuYuHSL} 
\end{eqnarray} 
reproduces with high accuracy the quantum Monte Carlo simulations\cite 
{Johnston}, as long as the ratio $J_{\parallel }/J_{\perp }$ does not exceed 
0.1. The definition ${\cal B}=J_{\perp }/(k_BT)$ is utilized in (\ref 
{GuYuHSL}). We compare our result of the susceptibility with (\ref{GuYuHSL}) 
of the HSL in Fig.\ref{comp0.1}. For more clarity of unit, we use the 
susceptibility scaled by the Curie constant. The comparison for  
$J_{\parallel }/J_{\perp }=0.1$ shows that our result is in a very good 
agreement with that of HSL, the coincidence will be even better for  
$J_{\parallel }/J_{\perp }<0.1$. 
 
\section{Summary} 
 
We have studied the quantum phase transitions, the magnetizations and 
susceptibility of an integrable spin ladder (ISL) by proposing a new way in 
deciding the system parameters to build closer connection of the ISL with 
the real compounds. The system parameters are chosen to make the extra 
interactions of the ISL relative to the Heisenberg spin ladder (HSL) vanish 
in the expectation of the ground state, especially for the gapped phase and 
fully-polarized phase. The analytic critical fields of such an ISL have the 
same leading terms as those of the experimental observations and the HSL:  
$g\mu _BH_{c1}\approx J_{\perp }-J_{\parallel }$ and $g\mu _BH_{c2}\approx 
J_{\perp }+2J_{\parallel }$, accompanied by the symmetric magnetization 
inflection point with the location analytically at $g\mu _BH_{IP}=J_{\perp 
}+J_{\parallel }/2$ (another result $g\mu _BH_{IP}=J_{\perp }+J_{\parallel 
}/2+dJ_{\parallel }/2$). A variety of comparisons have been made with other 
numerical results and the experiments. The magnetizations coincide with the 
finite-site exact diagonalization of the HSL for the GS, and with the TMRG 
numerical result for full temperatures. For experiments, the magnetizations 
agree well with the compounds (5IAP)$_2$CuBr$_4\cdot $2H$_2$O,  
Cu$_2$(C$_5$H$_{12}$N$_2$)$_2$Cl$_4$ and (C$_5$H$_{12}$N)$_2$CuBr$_4$,  
with the leg and 
rung coupling ratio $J_{\parallel }/J_{\perp }$ around 0.1, 0.2 and 0.3, 
respectively. The susceptibility with the clarified unit also coincides with 
the strongly-coupled compound as well as perturbation theory. The 
coincidence in the magnetization and susceptibility also makes it possible 
to discuss the specific heat for the whole process of the field application, 
we shall present the discussion in another paper. Our method may also be 
helpful for other integrable models in bridging to the experiments and real 
compounds. 
 
\section{Acknowledgments} 
 
ZJY thanks FAPERJ and FAPERGS for financial support. IR thanks PRONEX and 
CNPq. HQZ thanks Australian Research Council for financial support. 
ZJY also thanks Angela Foerster, Xi-Wen Guan, Murray.T. Batchelor, 
Bin Chen and Norman Oelkers for helpful discussions.

%\begin{thebibliography}{} 
 
%\end{thebibliography} 
 
\end{multicols}

%%%%%%%%%%%%%%%%%%%%%%%%%%%%%%%%%%%% Figure 1  
%%%%%%%%%%%%%%%%%%%%%%%%%%%%%%%%% end of Figure 1  
%%%%%%%%%%%%%%%%%%%%%%%%%%%%%%%%%%%% Figure 2  
%%%%%%%%%%%%%%%%%%%%%%%%%%%%%%%%% end of Figure 2  
%%%%%%%%%%%%%%%%%%%%%%%%%%%%%%%%%%%% Figure 3  
%%%%%%%%%%%%%%%%%%%%%%%%%%%%%%%%% end of Figure 3  
%%%%%%%%%%%%%%%%%%%%%%%%%%%%%%%%%%%% Figure 4  
%%%%%%%%%%%%%%%%%%%%%%%%%%%%%%%%% end of Figure 4  
%%%%%%%%%%%%%%%%%%%%%%%%%%%%%%%%%%%% Figure 5  
%%%%%%%%%%%%%%%%%%%%%%%%%%%%%%%%% end of Figure 5 
%%%%%%%%%%%%%%%%%%%%%%%%%%%%%%%%%%%% Figure 6  
%%%%%%%%%%%%%%%%%%%%%%%%%%%%%%%%% end of Figure 6 
%%%%%%%%%%%%%%%%%%%%%%%%%%%%%%%%%%%% Figure 7  
%%%%%%%%%%%%%%%%%%%%%%%%%%%%%%%%% end of Figure 7 
%%%%%%%%%%%%%%%%%%%%%%%%%%%%%%%%%%%% Figure 8  
%%%%%%%%%%%%%%%%%%%%%%%%%%%%%%%%% end of Figure 8  


\begin{references}
\bibitem{HaywardGap}  As the value of the gap $\Delta $ is not provided in 
Ref.\cite{Hayward} due to its variation for the presented various couplings, 
we take the exact second critical point $g\mu _BH_{c2}=J_{\perp 
}+2J_{\parallel }$ as a reference instead, marked by the symbol ``*''.
\end{references}
\end{document}